\documentclass[aps,prl,twocolumn,floatfix,superscriptaddress]{revtex4-2}

\usepackage{amsmath, amssymb}
\usepackage{enumitem}
\usepackage{mathtools}
\usepackage{lipsum}
\usepackage{times}
\usepackage{graphicx}
\usepackage{wasysym}
\usepackage{bm}
\usepackage[hidelinks]{hyperref}
\usepackage{xspace}
\usepackage{siunitx}
\usepackage{xcolor}
\usepackage{soul}
\usepackage{braket}
\usepackage[normalem]{ulem}
\usepackage[T1]{fontenc}

\definecolor{myColor}{rgb}{0,0,0.8}
\definecolor{myciteColor}{rgb}{0.39,0.7,0.89}
\hypersetup{colorlinks=true, linkcolor=myColor, filecolor=myColor, urlcolor=myColor,citecolor=myColor, urlcolor=myColor}
\graphicspath{{./}}

\DeclareSIUnit{\nK}{\nano\kelvin}
\DeclareSIUnit{\aB}{\emph{a}_0}
\DeclareSIUnit{\G}{G}

\renewcommand{\figurename}[1]{Fig.~}
\newcommand{\kB}{k_{\text{B}}}

\DeclareSIUnit\litre{l}
\newcommand{\IAP}{Institut für Angewandte Physik, Universität Bonn, Wegelerstrasse 8, 53115 Bonn, Germany}
\newcommand{\KIP}{Kirchhoff-Institut für Physik, Universität Heidelberg, Im Neuenheimer Feld 225a, 69120 Heidelberg, Germany}

\begin{document}

\title{Thermodynamics and State Preparation in a Two-State System of Light}

\author{Christian Kurtscheid}
\altaffiliation[Present address: ]{Fraunhofer-Institut für Hochfrequenzphysik und Radartechnik FHR, Fraunhoferstr. 20, 53343 Wachtberg, Germany}
\affiliation{\IAP}
\author{Andreas Redmann}
\email[]{redmann@iap.uni-bonn.de}
\affiliation{\IAP}
\author{Frank Vewinger}
\affiliation{\IAP}
\author{Julian Schmitt}
\affiliation{\IAP}
\affiliation{\KIP}
\author{Martin Weitz}
\email[]{martin.weitz@uni-bonn.de}
\affiliation{\IAP}

\begin{abstract}
The coupling of two-level quantum systems to the thermal environment is a fundamental problem, with applications of such usually single-particle or fermionic systems ranging from qubit state preparation to spin models. The present work studies the elementary problem of the thermodynamics of an ensemble of bosons populating a two-level system. Using an optical dye microcavity platform, we thermalize photons in a two-mode system at conditions of tunable chemical potential, demonstrating the statistical mechanical problem of $N$ bosons populating a two-level system, coupled to a heat bath. Under pulsed excitation, we observe Josephson oscillations between the two quantum states, which verifies the possibility for coherent manipulation. In contrast, under stationary conditions the thermalization of the two-mode system is observed, arising from radiative coupling of photons to the dye. As the energetic splitting between eigenstates is two orders of magnitude smaller than thermal energy, at low occupations an almost equal distribution of the modes occupation is observed, as expected from Boltzmann statistics. For larger occupation, we observe efficient population of the ground state and saturation of the upper level at high filling, expected from quantum statistics. Our experiment holds promise for state preparation in quantum technologies as well as for quantum thermodynamics studies.
\end{abstract}

\maketitle

Thermodynamic properties of light are well studied for gaseous systems - photon gases - possessing a continuum of states, with the most common example for this being blackbody radiation \cite{Huang:1987}. In this three-dimensional textbook system the radiant power, following the Stefan Boltzmann law $P \propto T^4$, is a fixed function of temperature $T$, and at low temperatures photons vanish in the system walls. More recently also lower-dimensional systems have been studied, such as the two-dimensional photon gases realized in dye-filled microcavities for which the ‘frozen’ dimension gives rise to a low-frequency cutoff and the chemical potential becomes tuneable, much as in a gas of material particles \cite{Klaers:2010b}. Such systems have demonstrated to allow for photon Bose-Einstein condensation, leading to a macroscopic ground state population in a non-trivial lowest energy mode. Bose-Einstein condensation has in different configurations been realized both in photon and polariton systems \cite{Klaers:2010, Marelic:2015, Greveling:2018, Vretenar:2021a, Schofield:2024, Pieczarka:2024, Kasprzak:2006, Balili:2007, Deng:2010, Bloch:2022}, furthermore the variability of these platforms has allowed to apply variable potentials. This has for example allowed for the observation of Josephson physics with photons or polaritons \cite{Vretenar:2021b, Lagoudakis:2010, Abbarchi:2013}. Thermalization effects of photons have also been observed in optical multimode fibers \cite{Aschieri:2011, Baudin:2020, Pourbeyram:2022, Ferraro:2024}. In other work, upon thermalization of photons in a double-well system superimposed by a harmonic trap potential as to provide the required density of states of the higher lying mode continuum a Bose-Einstein condensate in a split state of light has been realized \cite{Kurtscheid:2019}.

In the extreme opposite limit to the continuous systems described above, two-state systems are similarly ubiquitous in quantum physics. The most well-known example probably is a spin \textonehalf-system, such as a proton in a magnetic field, where energy levels are separated by the Zeeman splitting. Near resonant AC fields can drive transitions between these levels, while initial state preparation in liquid and solid state systems rely on the imbalance of equilibrium populations of the two quantum states at ambient temperature \cite{Abragam:1983, Levitt:2013, Faghihi:2017}. The obtainable signal in NMR applications crucially depends on the fidelity of state preparation, which here is well described by applying a Boltzmann distribution $e^{-\Delta E/ \kB T}$ , where $\Delta E$ denotes the Zeeman splitting and $T$ temperature, which at typical fields of 4 Tesla, corresponding to $\Delta E \simeq h \cdot \SI{200}{\mega\hertz}$, yields a $\sim 10^{-5}$ excess population of the low energy ground state at room temperature. Two-state systems, usually effectively fermionic, coupled to a bath are commonly examined in the context of the spin-boson model \cite{DeRaedt:1984, Carmeli:1985, Weiss:2008, Kockum:2019}, and coupled spin systems can be described e.g. by the Ising model \cite{Sachdev:2011}. For bosonic cold atom systems, Bose-Einstein condensates have been adiabatically loaded from harmonic into double-well traps \cite{Albiez:2005}, which has e.g. allowed to prepare non-classical many-body states \cite{Esteve:2008}.

In this work, we examine thermodynamic properties of photons populating a two-state system with a freely tuneable chemical potential. Photons are thermalized by radiative coupling to a bath of dye molecules. The two-state system is provided by two modes of a microstructured dye-filled optical microcavity, where the coherent coupling is verified by the observation of Josephson oscillations. The energetic splitting between the two eigenmodes is two orders of magnitude below the thermal energy, and under equilibrium conditions at low filling we observe an almost equal distribution of eigenstate populations as expected from Boltzmann statistics, which corresponds to the situation of state preparation in room temperature NMR experiments. At high occupation numbers, however, despite the small energetic splitting more than 90\% of the measured emission arises from population of the low energy eigenmode, as well understood from bosonic stimulation imposed by quantum statistics. Our results are in good agreement with the predictions of the elementary statistical mechanics model of $N$ bosons occupying a two-state system in thermal equilibrium. 

\begin{figure*}[t]
    \centering
    \includegraphics[width=1.0\textwidth]{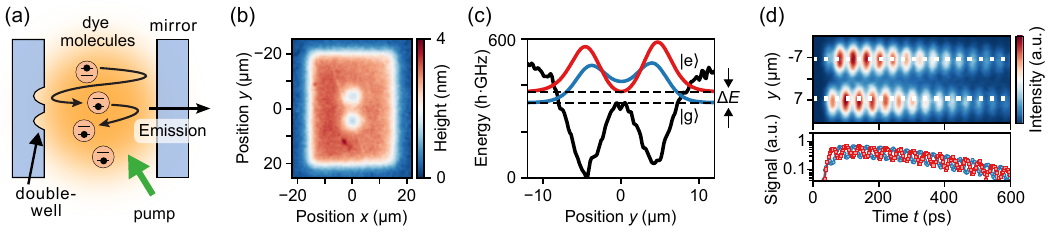}
\caption{(a)~Photons are confined in a dye-solution filled optical microresonator consisting of one plane and one mirror with a laterally microstructured reflective surface profile. While trapped in the microresonator, photons thermalize with the dye by repeated absorption re-emission processes. (b)~Height profile of the microstructured cavity mirror. (c)~The black solid line gives a cut of the corresponding expected potential for cavity photons along the axis of the double well, realizing a two-level system with ground ($\ket{\mathrm{g}}$) and excited ($\ket{\mathrm{e}}$) states, see the dashed black lines for the (calculated) position of energy levels. The blue and red lines give the spatial variation of the probability densities of the two optical eigenstates along this axis respectively. (d)~(Top panel) Tunneling dynamics of photons in a double-well structure where the oscillations at the used pump pulse duration can be resolved. (Bottom panel) The connected blue circles and red squares give the time dependence of the population signals in the upper and lower sites respectively at $x=\pm \SI{7}{\micro\meter}$ (dashed lines, top panel).}
    \label{fig:1}
\end{figure*}

A schematic of our experimental scheme is shown in Fig.~\ref{fig:1}(a). The used setup is a modified version of an apparatus used in earlier works~\cite{Kurtscheid:2019, Kurtscheid:2020} (see End Matter for details). Photons are confined in a dye-solution filled microcavity with mirrors spaced in the wavelength regime ($D_0 \simeq \SI{1.8}{\micro\meter}$). The small spacing introduces a large mode spacing that is comparable to the emission width of the dye, such that effectively only a single longitudinal cavity mode ($q=9$) is populated, which yields an effective low-frequency cutoff at $hc / \lambda_\mathrm{cutoff} \simeq \SI{2.1}{\electronvolt}$ photon energy, where $\lambda_\mathrm{cutoff} = 2 n_\mathrm{r} D_0 /q$  (typically \SI{589}{\nano\meter}) denotes the cutoff wavelength, and $n_\mathrm{r} = 1.44$ is the refractive index. Moreover, the photon dispersion becomes quadratic, as for a massive particle. To generate a two-state system, a double well potential is imprinted for the photon gas by microstructuring of the surface profile of one of the cavity mirrors, which leads to a position-dependent permanent lifting of the reflecting dielectric coating~\cite{Kurtscheid:2020}. Fig.~\ref{fig:1}(b) shows a Mirau image of the created surface profile on the mirror, with two indents of $\simeq \SI{5}{\micro\meter}$ circular diameter and \SI{1.6}{\nano\meter} depth spaced by \SI{8.6}{\micro\meter} on the otherwise plane mirror surface. Within the microcavity at the lateral position of the indents, longer wavelength, i.e. lower energetic, photons fulfill the boundary conditions, i.e. we create a locally attractive potential. One can show that the photon gas in the resonator is formally equivalent to a two-dimensional gas of massive particles with effective mass $m_\mathrm{eff} = h / (c \cdot \lambda_\mathrm{cutoff})$ subject to a potential $V(x,y) = m_\mathrm{eff} c^2 \cdot h(x,y) / D_0$, where $h(x,y)$ denotes the local height. The solid line in Fig.~\ref{fig:1}(c) gives the variation of the generated potential for cavity photons along the axis of the double well. In the two-site potential, two eigenstates are trapped, as well described by the symmetric and antisymmetric superpositions of localized site wavefunctions, see the dashed black lines for the calculated eigenenergies and the blue and red solid lines for the probability densities of the eigenstates. This realizes a two-state system for cavity photons, with a splitting of $\Delta E = h \cdot \SI{50}{\giga\hertz}$ for the structure shown in Fig.~\ref{fig:1}(b).

In initial experiments, we have studied Josephson oscillations in the two-state system. For this, the dye microcavity was pumped by pulsed laser irradiation, of typical pulse length \SI{20}{\pico\second} at \SI{532}{\nano\meter} wavelength. To allow for a temporal resolving of oscillations triggered by the used pulses, for this measurement a structure with a smaller energy difference between eigenmodes, resulting in smaller oscillation frequencies as realized by using a larger distance, of \SI{12}{\micro\meter}, between the microsites, was used. The pump beam was tightly focused (\SI{10}{\micro\meter} beam diameter), and directed to one of the microsites of the double-well system. We note that for this measurement due to the strong pumping the intracavity intensity (typical photon numbers: $10^6$ in the two modes) was too high to allow for significant thermalization effects of cavity photons \cite{Hesten:2018}, such that long-lasting wavepacket dynamics as set from the initial state occurred (see End Matter for details). Typical experimental data, as observed with a streak camera, is shown in the top panel of Fig.~\ref{fig:1}(d) as a function of both time and position along the axis of the double well. We observe oscillations of the photon wavepackets between the trapping sites, see also the bottom panel for a plot giving the temporal evolution of populations. The decay at long times ($\sim \SI{190}{\pico\second}$ time constant) arises from the finite lifetime of cavity photons due to mainly mirror losses. The oscillation occurs at a frequency that corresponds to the energy difference of the two eigenstates in frequency units, which for the here used structure equals \SI{24}{\giga\hertz}. The measurements show that coherent manipulation between eigenstates of such a two-state system is possible.

In the next step, we have studied thermalization of photons in the double-well potential. For a bosonic gas with freely tunable chemical potential $\mu$, the expected populations $n_i$ of eigenstates with energies $E_i$ ($i= \{\mathrm{g},\mathrm{e}\}$) are described by the Bose-Einstein distribution $n_i = g_i / (\exp[(E_i - \mu)/\kB T] - 1)$, where $g_i$ ($=2$ here accounting for photon polarization) denotes the mode degeneracy, $N = n_\mathrm{g} + n_\mathrm{e}$ is the total population, and we set $E_\mathrm{g} = 0$ to the cutoff energy and $E_\mathrm{e} = \Delta E$. It is instructive to discuss the respective populations in the here relevant limit of the energetic splitting $\Delta E$ being clearly below the thermal energy. We then have both $n_\mathrm{g} \gg 1$ and $n_\mathrm{e} \gg 1$ such that the populations are well approximated by $n_\mathrm{g} = -2 \kB T/\mu$, $n_\mathrm{e} = 2 \kB T / (\Delta E - \mu)$. The chemical potential is determined through photon number conservation by $N = 2 \kB T (1/(\Delta E - \mu) - 1/\mu)$, which yields a quadratic equation for $\mu$ that can be solved analytically, so that a closed expression for the respective population is obtained: 
\begin{align}
    n_{\mathrm{e,g}} = \frac{N}{2} \pm \frac{2 \kB T}{\Delta E} \mp \sqrt{\frac{4 \kB^2 T^2}{\Delta E^2} + \frac{N^2}{4}} \, .
    \label{eq:1}
\end{align}
At particle numbers $N$ much smaller than a characteristic photon number $N_\mathrm{c} = 2 \kB T / \Delta E$, the result for the population distribution in this approximation with $n_\mathrm{e,g} = N/2$ yields an equalized distribution between the two quantum states as expected from Boltzmann statistics. On the other hand, in the limit of a particle number $N \gg N_\mathrm{c}$, we obtain $n_\mathrm{e} = N_\mathrm{c}$, and $n_\mathrm{g} = N - N_\mathrm{c}$, meaning that we expect that the upper state saturates at a maximum particle number of $N_\mathrm{c} \simeq 250$ here (for $\Delta E \simeq h \cdot \SI{50}{\giga\hertz}$), and the ground state of the two-level system takes all remaining population. It should be pointed out that the corresponding transition between these two limits which occurs near $N_\mathrm{c}$ is not sharp as for Bose-Einstein condensation, but is a smooth crossover, as well understood from that the two-level system naturally is rather the opposite of a thermodynamic limit.

\begin{figure}[t]
    \centering
    \includegraphics[width=1.0\columnwidth]{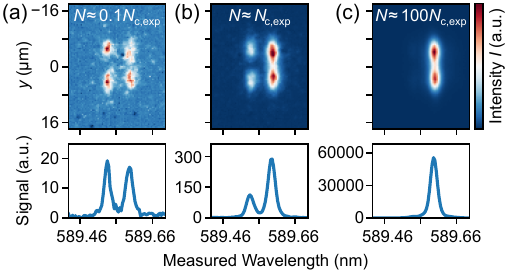}
\caption{The top panel gives the measured wavelength as a function of the transverse position along the axis of the double well ($y$-axis) for three different values of the photon number $N$ in the microcavity: (a)~for $N \simeq 0.1 \cdot N_\mathrm{c,exp}$, (b)~for $N \simeq N_\mathrm{c,exp}$, and (c)~for $N \simeq 100 \cdot N_\mathrm{c,exp}$, where $N_\mathrm{c,exp}$ denotes the characteristic photon number. The two peaks visible on the right hand side at lower photon energy (higher wavelength) correspond to the signal from photons in the (symmetric) ground state, the peaks on the left hand side to the (antisymmetric) excited state of the two-level system. The bottom panels show spectra obtained by integrating the data along the $y$-axis. While at low photon numbers (a) the population distribution between sublevels is comparable, upon increasing the photon number (b,c) the relative population in the ground state, see the right hand side peak, clearly dominates. The scale of the color code in the top panels is normalized to the total photon number in each of the respective images (a-c).}
    \label{fig:2}
\end{figure}

To experimentally study thermalization of photons in the double-well potential, quasi-cw pumping of the dye microcavity was employed (see End Matter for details). At the used conditions (here up to \SI{100}{\milli\watt} pump power on a \SI{70}{\micro\meter} beam diameter was used) repeated absorption re-emission processes thermalize cavity photons to the (rovibrational) temperature of the dye, which is at room temperature. Given that the cutoff energy is clearly above the thermal energy $\kB T$ ($\simeq 1/\SI{40}{\electronvolt}$) direct creation of photons by thermal emission is suppressed. Different than in blackbody radiation the photon number is thus conserved and the chemical potential is freely tuneable, as applies for a bosonic gas of massive particles, e.g. ultracold atoms, see also earlier work testing for the thermalization of photons in such short dye microcavities \cite{Klaers:2010b}.

\begin{figure}[t]
    \centering
    \includegraphics[width=1.0\columnwidth]{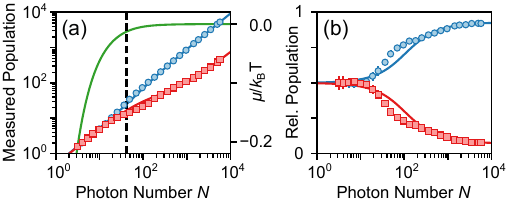}
\caption{(a)~Variation of the observed populations in upper (red squares) and lower (blue dots) states of the two-level system with total photon number. The solid lines are theory fits, accounting for a finite instrumental spectrometer resolution. The vertical dashed line gives the position of the characteristic photon number $N_\mathrm{c,exp}$, and the green solid line gives the chemical potential (see main text). (b)~Corresponding variation of the measured relative photon populations in upper (red squares) and lower (blue dots) states, along with fits. The data shows that clearly below the characteristic photon number the populations in upper and lower levels are almost equally distributed, while at high photon numbers a significantly enhanced population in the lower energetic ground state is observed. Data points are the mean values of $\sim 150$ experimental realizations per binned photon number and error bars show the statistical standard deviations.}
    \label{fig:3}
\end{figure}

To analyze the population distribution of the optical two-state system, the cavity emission was directed over a homebuilt high-resolution slitless spectrometer and then imaged onto an intensified camera. The upper panels of figs.~\ref{fig:2}(a)-(c) in a color-code representation give typical recorded spectrally dispersed images with the position along the axis of the double-well potential (vertical axis) resolved, and the bottom panels corresponding spectra, as obtained by binning along the vertical axis. In the three spectra, the respective peak on the right hand side at the cutoff wavelength ($\lambda_\mathrm{cutoff} \simeq \SI{589.6}{\nano\meter}$) stems from photons populating the ground state mode $\ket{\mathrm{g}}$ of the two-level system, while the peak at lower wavelength (higher photon energy) is the signal in the upper state $\ket{\mathrm{e}}$. The asymmetric nature of the latter state gives rise to the visible larger spatial separation of the intensity maxima along the vertical direction as compared to the ground state, which is the symmetric linear combination of localized wavefunctions. While at low total photon numbers, we observe an essentially equal distribution of photons between the eigenstates (Fig.~\ref{fig:2}(a)), at larger total photon numbers the relative contribution of the ground state population is enhanced and for large photon numbers (Fig.~\ref{fig:2}(c)) the ground state population clearly dominates.

Our experimental result for the variation of the observed photon number in the ground and excited states of the bosonic two-state system, as deduced from the spectrometer signal, versus the total photon number is shown in Fig.~\ref{fig:3}(a), and Fig.~\ref{fig:3}(b) gives the corresponding relative occupations of the two quantum states. The essentially equal population of the two modes observed in the limit of small photon numbers is well in agreement with theory. Here Boltzmann statistics is expected to well apply, from which an imbalance of $\Delta E / \kB T \simeq 0.008$ is expected, which is below the experimental uncertainties. For increased particle numbers well above 100 photons the upper state population shows signatures of saturation, and the ground mode takes most of the additional photons, see e.g. the blue dots and the red squares giving the relative populations in lower and upper quantum states in Fig.~\ref{fig:3}(b). Given that the energetic splitting, which in temperature units is of magnitude \SI{2.4}{\kelvin}, is much smaller than the thermal energy ($k_B \times \SI{300}{\kelvin}$), we attribute the clearly observed enhanced population of the lower energetic state to the bosonic stimulation from the manybody population in the two-level system. At high photon numbers (for $N \gg N_\mathrm{c}$), the measured population in the ground state reaches 93\% of the total photon number, which means that we observe an imperfect saturation for the upper level, which here takes the role of the thermal cloud in BEC experiments. The origin of this incomplete saturation of the observed signal is attributed to be caused mainly by the finite resolution of the used spectrometer of \SI{20}{\giga\hertz}, which instrumentally limits our ability to spectrally distinguish the nearby peaks (\SI{50}{\giga\hertz} separation). The experimental data can well be modeled with an ansatz $n_\mathrm{e,spec} = C n_\mathrm{e}^\mathrm{{eq}} + (1-C) n_\mathrm{g}^\mathrm{eq}$, $n_\mathrm{g,spec} = (1-C) n_\mathrm{e}^\mathrm{eq} + C n_\mathrm{g}^\mathrm{eq}$, where $n_\mathrm{e,g}^\mathrm{eq} = \frac{N}{2} \pm N_\mathrm{c,exp}^2 \mp \sqrt{N_\mathrm{c,exp}^2 + N^2/4}$ as the distributions (see eq.~\eqref{eq:1}) for a room temperature (\SI{300}{\kelvin}) Bose-Einstein distribution in the here relevant limit $\Delta E \ll \kB T$, where $N_\mathrm{c,exp}$ as the experimental value for the characteristic photon number and the parameter $C$ modeling the relative contrast of populations, derived from the spectrometer data, assuming finite resolving power, were taken as free fit parameters. For the data of Fig.~\ref{fig:3}, the fit yields $C\simeq 0.93$ which is close to the ideal case $C=1$. The fitted value for the characteristic photon number $N_\mathrm{c,exp} \approx 48$ is below the expected value of $N_\mathrm{c} \simeq 250$, indicating that the above described idealized loss-free model has to be augmented to capture the driven-dissipative nature of the experimental system. We nevertheless observe that the main features of the measured data are well captured by the theory model.

We emphasize, that other than in the effect of Bose-Einstein condensation, indeed all system states are macroscopically occupied and there is no true thermal cloud, with the upper state population being clearly above unity. The observed saturation behavior above a characteristic photon number clearly distinguishes itself from expectations in laser models for varying pump \cite{Siegman:1986}. The green solid line in Fig.~\ref{fig:3}(a) shows the expected variation of the chemical potential, which is strongly negative (with $|\mu| \gg \Delta E$) at low photon numbers and approaches zero ($|\mu| \ll \Delta E$) in the limit of $N \gg N_\mathrm{c,exp}$, which qualitatively resembles the situation in Bose-Einstein condensates. While in the case of Bose-Einstein condensation, where a continuum of modes with energy bandwidth above the thermal energy results in the presence of a phase transition to a macroscopically occupied quantum state, here a system with a single additional state (beyond the ground state) of excitation energy below the thermal energy is considered, upon which a crossover to a phase with above 90\% ground state population is observed.

\begin{figure}[t]
    \centering
    \includegraphics[width=1.0\columnwidth]{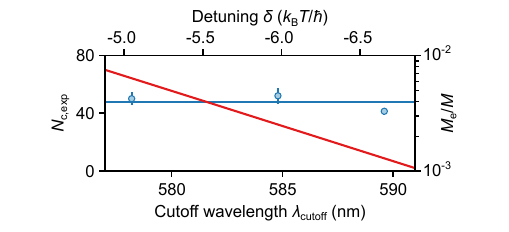}
\caption{Variation of the characteristic photon number $N_\mathrm{c,exp}$ (blue dots) on the low-frequency cutoff wavelength $\lambda_\mathrm{cutoff}$ of the microcavity. The data points are fitted with a constant term (see the blue line), which corresponds to expectations from thermodynamic theory. The red solid line shows the corresponding calculated dependence of the number of dye molecular excitations $M_\mathrm{e}$, in units of the total molecule number $M$, on the cutoff wavelength. Fitted values for the measured contrast of populations are $C$ = \SI{0.912 \pm 0.001}{}, \SI{0.921 \pm 0.001}{}, and \SI{0.928 \pm 0.001}{} for $\lambda_\mathrm{cutoff}$ = \SI{578.2}{\nano\meter}, \SI{584.8}{\nano\meter}, and \SI{589.6}{\nano\meter} respectively.}
    \label{fig:4}
\end{figure}

In further measurements, we have recorded data at different values of the cutoff wavelength, which via the condition of chemical equilibrium between photons and dye electronic excitations acting as a reservoir tunes the number of dye electronic excitations, as well as the dye absorption which determines the thermalization time \cite{Klaers:2012, Schmitt:2015}. Still, quite comparable results have been obtained for the measured two-level populations as in the data of Figs.~\ref{fig:2} and \ref{fig:3}. The data points in Fig.~\ref{fig:4} show the variation of the fitted value of the characteristic photon number $N_\mathrm{c,exp}$ on the cutoff wavelength $\lambda_\mathrm{cutoff}$ for the corresponding measurements, and the red line gives the expected dependence of the number of dye excitations $M_\mathrm{e}$, as obtained from $M_\mathrm{e} / M = 1/(\exp{(-\hbar \delta / \kB T)} + 1)$, which is valid for $N \gtrsim N_\mathrm{c}$. Here we have $\delta = \omega_\mathrm{cutoff} - \omega_\mathrm{zpl}$, with $\omega_\mathrm{cutoff} = 2 \pi c / \lambda_\mathrm{cutoff}$, $\omega_\mathrm{zpl} = 2 \pi c / \lambda_\mathrm{zpl}$, where $\lambda_\mathrm{zpl} = \SI{545}{\nano\meter}$ is the zero-phonon line of the dye, and $M \simeq 10^8$ is the total number of dye molecules. Despite the variation of system parameters, the spread of the obtained data for the characteristic photon number for the different measurements is comparable to the experimental uncertainties, such that this data can be well fitted with a constant (blue line), as expected from theory. In other words, the position of the crossover to a phase with macroscopic ground state occupation within experimental uncertainties is independent of the reservoir size, which gives further clear evidence that the observed effects are governed by the thermodynamics of the two-mode system of light. 

To conclude, we report on the thermalization of a two-state system of light with non-vanishing chemical potential. Despite that the energetic splitting between eigenstates - and thus also the total energetic bandwidth of relevant states - is two orders of magnitude below the thermal energy, we observe efficient population of the lower energetic state of the two-level system. The results are in good agreement with quantum statistical mechanics theory of the bosonic $N$-particle system.

Prospects of the observed effects include state preparation in quantum engineering applications based on coupled-macroscopic quantum systems, as $N$-qubit quantum rings~\cite{Xue:2021, Barrat:2024}, novel sensing applications~\cite{Bennet:2020}, and for quantum key distribution~\cite{Da:2019}. Technically, such applications could benefit from exploring the integration of material-filled optical microcavities into waveguide arrays~\cite{Lenzini:2018}. In the high photon number limit, the demonstrated two-state system acts as a non-unitary beamsplitter, which in the idealized, lossless case is capable of combining two input channels into a single output channel, as is thermodynamically possible due to coupling with a reservoir. Interestingly, this is a concept that in principle even allows for the power scaling of optical sources without the need for phase locking, which is required when combing with conventional, unitary beam splitters~\cite{Linslal:2022}. Further, with the use of birefringent or optically active materials in the cavity, prospects for thermodynamic polarisation control of emitters arise. Other perspectives include studies of the spin-boson model~\cite{Weiss:2008} as well as quantum thermodynamics~\cite{Tude:2024, Williamson:2024}.


\begin{acknowledgments}
We acknowledge support from the DFG within SFB/TR 185 (277625399), the Cluster of Excellence ML4Q (390534769), and the DLR project BESQ with funds provided by the BMWK (50 WM2240). J.S. acknowledges support by the EU (ERC, TopoGrand, 101040409). 
\end{acknowledgments}

The data that support the findings of this article are openly available~\cite{Zenodo:2025}.

\section{End Matter}
\textit{Preparation of the double-well microstructure} -- To implement a double-well potential for photons in a dye-microcavity environment, one of the two cavity mirrors is microstructured in terms of its reflecting surface profile. For this, a static mirror surface structuring technique is used~\cite{Kurtscheid:2020}. This method is employed for ultrahigh reflectivity dielectric mirrors equipped with an additional \SI{30}{\nano\meter} thick amorphous silicon layer between the substrate and the dielectric coating stack consisting of alternating amorphous tantalum pentoxide and silicon oxide stacks. The silicon layer is absorptive for visible laser light, which is a key feature to enable mirror surface structuring. For structuring, which is performed prior to operation of the microcavity, a laser beam near \SI{532}{\nano\meter} wavelength focused to $\sim\SI{1.8}{\micro\meter}$ diameter and with an acousto-optic modulator chopped to $\sim\SI{10}{\nano\second}$ long pulses is directed through the back side of the mirror to the absorptive layer. The laser-induced heating is observed to lead to a local lifting of the reflecting mirror surface, as attributed to a controlled delamination of the dielectric coatings from the substrate due to heat-induced stress as well as pore formation in the lowest layers of the coatings~\cite{Kurtscheid:2020, Vretenar:2023}. By lateral steering of the heating laser beam focus by means of a galvanometer-scanner and controlling the beam intensity with the acousto-optic modulator, a variable heat pattern is induced to lift the reflecting mirror surface to a desired target profile. The mirror surface spatial profile is monitored by means of white light, phase shifting optical interferometry, as realised using a Mirau-objective with an in-built reference path to interfere light reflected from the mirror surface. The measurement is performed by piezo-electrically moving the mirror sample along the optical axis over a distance of $\sim\SI{1}{\micro\meter}$ while a camera continuously records a spatially resolved interference pattern. From an analysis of a series of corresponding interference patterns, the mirror surface profile is determined. To generate the desired mirror profiles for variable potentials in our optical microcavity, an iterative procedure is applied in which we repeatedly alter between the laser writing procedure and optical testing of the surface structure. Using the iterative procedure, variable mirror structures with typical mean deviations between the target profile and the desired structure of $\sim\SI{1}{\angstrom}$ can be obtained.

\textit{Dye microcavity experimental apparatus} -- The used dye microcavity consists of both a structured as well as a usual unstructured plane dielectric mirror spaced by $D_0 \simeq \SI{1.8}{\micro\meter}$, filled with rhodamine 6G dye solution dissolved in ethylene glycol at $\SI{1}{\milli\mol\per\litre}$ concentration. The mirror reflectivity is $\gtrapprox 99.997\%$, with the structuring for the used elevation heights not seemingly affecting the reflectivity. The dye microcavity is pumped at about $\SI{45}{\degree}$ angle to the optical axis with a laser beam near \SI{532}{\nano\meter} wavelength and using a pump beam diameter of $\sim \SI{70}{\micro\meter}$ (or $\sim \SI{10}{\micro\meter}$ diameter for the Josephson oscillation experiments). For analysis, the fluorescence emitted from the dye microcavity was collected through the unstructured mirror, and both spatially and spectrally resolved analysed (spatially and temporally resolved for the Josephson oscillation work respectively), as described in the main text.

\textit{Josephson dynamics} -- The wavepacket dynamics observed when directing a short, intense pump pulse (typical intracavity photon numbers: $10^6$) to one of the double-well microsites arises from the beating of energy eigenstates. For a simple model, let $|\mathrm{L}\rangle$ and $|\mathrm{R}\rangle$ represent the localized eigenstates of the individual wells. The eigenstates of the coupled double-well system, with energies $E=0$ and $E=\Delta E$, can then be written as $|\mathrm{g}\rangle=1/\sqrt{2} \, (|\mathrm{L}\rangle + |\mathrm{R}\rangle)$ and $|\mathrm{e}\rangle = 1/\sqrt{2} \, (|\mathrm{L}\rangle - |\mathrm{R}\rangle)$ respectively. Oscillatory dynamics emerges when the energy eigenstates are prepared in a coherent superposition with a well-defined initial phase, as set by the short pump pulse. The temporal evolution can then be described in a wavefunction description as $|\varphi(t)\rangle = \alpha(0)|\mathrm{g}\rangle + \beta(0) \exp{(-i\Delta E t /\hbar)}|\mathrm{e}\rangle$, with initial amplitudes $\alpha(0)$ and $\beta(0)$. For an initial state $|\varphi(0)\rangle =
|\mathrm{L}\rangle$, as prepared by the pump pulse at $t=0$, we have $\alpha(0)=\beta(0)=1/\sqrt{2}$, such that the expected population dynamics is the usual Josephson dynamics $|\langle \mathrm{L}, \mathrm{R}|\varphi (t)\rangle|^2=(1/2)(1\pm \cos(\Delta E t /\hbar))$. The experimental signal (Fig.~\ref{fig:1}(d), main text) well shows a corresponding coherent oscillation, though the simple model does not include coupling to the dye molecules as well as photon loss. To resolve the coherent dynamics, the system was prepared under experimental conditions far from equilibrium (strong and temporally short excitation pulse), leading to non-vanishing off-diagonal elements in the photon density matrix. The opposite limit of weak excitation, where thermalization leads to a decay of the off-diagonal coherences (see e.g.:~\cite{Schmitt:2015}), was in the present work not studied using the pulsed laser source but by employing a lower intensity chopped cw pump laser, described in the next section.

\textit{Thermalization of photons in the microcavity} -- For the experiments investigating thermalization of the two-state system, a cw-laser was used for pumping and the, as compared to the distance between microsites, used large diameter of the pump beam ensured that the double-well structure was well irradiated spatially homogeneously. The pump beam was acoustooptically chopped into $\sim \SI{500}{\nano\second}$ pulses to suppress photo-bleaching and triplet pumping of the dye, and we use a $\SI{50}{\hertz}$ repetition rate. The length of the pump pulses is at least two orders of magnitude longer than thermalization, which proceeds at a timescale of the rhodamine excited state spontaneous lifetime ($\sim \SI{4.1}{\nano\second}$) or faster when stimulated processes come into play~\cite{Schmitt:2015, Kirton:2013}, such that the experiment can well be considered to operate quasi-cw. Photons confined in the microresonator are driven towards a thermalized state by multiple absorption and emission processes, which results in thermal contact with the dye solution. This stems from a thermalization of the rovibronic states of the dye molecules, as due to frequent collisions of solvent molecules of the dye, which occur on a time scale of $10 - \SI{100}{\femto\second}$, i.e. much faster than electronic processes in the dye. Correspondingly, photon absorption and emission can well be assumed to occur from thermally equilibrated rovibrational manifolds from lower and upper electronic states respectively. One can show that the spectral distributions of absorption $\alpha(\omega)$ and emission $f(\omega)$ are then linked by the Boltzmann-like Kennard-Stepanov frequency scaling $f(\omega) / \alpha(\omega) \propto \omega^3 \exp{(-\hbar \omega / \kB T)}$ (see, e.g.:~\cite{Lakowicz:2006}). By multiple absorption and emission processes, the thermalized distribution is then transferred to the spectral distribution of the photon gas as the photons many times cycle back and forth between resonator mirrors. This drives the system towards a Bose-Einstein distribution of cavity eigenmodes, see also earlier works~\cite{Klaers:2010b}. The chemical potential is freely tuneable, and can here experimentally be adjusted by the pump beam power. The tunneling time between the sites of the double-well structure is below the reabsorption time of the dye in the absorption re-emission cycles, such that we expect that the relevant eigenstates for the thermalization are those of the coupled double-well system. For the description of the thermalized state a density matrix formalism may be used, with $\rho_{\mathrm{gg}}\equiv n_\mathrm{g}$, $\rho_{\mathrm{aa}}\equiv n_\mathrm{a}$ as the populations of the eigenstates of the double-well system and $\rho_{\mathrm{ga}} = \rho_{\mathrm{ag}}^*$ as the off-diagonal elements. Due to the quasi-cw pumping used in these experiments, with the length of the pump pulses being four orders of magnitude longer than the oscillation time, no wavepacket dynamics can occur and here the off-diagonal density matrix elements vanish. The presence of a thermalized state can be tested for by analyzing the populations for different pump intensities (see main text). We note that the transition between wavepacket dynamics initiated by pulsed pump laser irradiation and thermalization has been confirmed in previous experiments with a single extended harmonic trapping potential for the photons in the dye-microcavity system~\cite{Schmitt:2015, Keeling:2016}, where it has been shown that upon the onset of thermalization wavepacket dynamics ceases, as expected. Further, in the presence of losses, the chemical potential is no longer perfectly spatially homogeneous~\cite{Keeling:2016}, which for the here studied two-level system can cause a shift of the characteristic photon number due to the remaining driven-dissipative nature.


\begin{thebibliography}{47}%
	\makeatletter
	\providecommand \@ifxundefined [1]{%
		\@ifx{#1\undefined}
	}%
	\providecommand \@ifnum [1]{%
		\ifnum #1\expandafter \@firstoftwo
		\else \expandafter \@secondoftwo
		\fi
	}%
	\providecommand \@ifx [1]{%
		\ifx #1\expandafter \@firstoftwo
		\else \expandafter \@secondoftwo
		\fi
	}%
	\providecommand \natexlab [1]{#1}%
	\providecommand \enquote  [1]{``#1''}%
	\providecommand \bibnamefont  [1]{#1}%
	\providecommand \bibfnamefont [1]{#1}%
	\providecommand \citenamefont [1]{#1}%
	\providecommand \href@noop [0]{\@secondoftwo}%
	\providecommand \href [0]{\begingroup \@sanitize@url \@href}%
	\providecommand \@href[1]{\@@startlink{#1}\@@href}%
	\providecommand \@@href[1]{\endgroup#1\@@endlink}%
	\providecommand \@sanitize@url [0]{\catcode `\\12\catcode `\$12\catcode
		`\&12\catcode `\#12\catcode `\^12\catcode `\_12\catcode `\%12\relax}%
	\providecommand \@@startlink[1]{}%
	\providecommand \@@endlink[0]{}%
	\providecommand \url  [0]{\begingroup\@sanitize@url \@url }%
	\providecommand \@url [1]{\endgroup\@href {#1}{\urlprefix }}%
	\providecommand \urlprefix  [0]{URL }%
	\providecommand \Eprint [0]{\href }%
	\providecommand \doibase [0]{https://doi.org/}%
	\providecommand \selectlanguage [0]{\@gobble}%
	\providecommand \bibinfo  [0]{\@secondoftwo}%
	\providecommand \bibfield  [0]{\@secondoftwo}%
	\providecommand \translation [1]{[#1]}%
	\providecommand \BibitemOpen [0]{}%
	\providecommand \bibitemStop [0]{}%
	\providecommand \bibitemNoStop [0]{.\EOS\space}%
	\providecommand \EOS [0]{\spacefactor3000\relax}%
	\providecommand \BibitemShut  [1]{\csname bibitem#1\endcsname}%
	\let\auto@bib@innerbib\@empty
	\bibitem [{\citenamefont {Huang}(1987)}]{Huang:1987}%
	\BibitemOpen
	\bibfield  {author} {\bibinfo {author} {\bibfnamefont {K.}~\bibnamefont
			{Huang}},\ }\href@noop {} {\emph {\bibinfo {title} {Statistical Mechanics}}}\
	(\bibinfo  {publisher} {Wiley},\ \bibinfo {address} {New York},\ \bibinfo
	{year} {1987})\BibitemShut {NoStop}%
	\bibitem [{\citenamefont {Klaers}\ \emph
		{et~al.}(2010{\natexlab{a}})\citenamefont {Klaers}, \citenamefont
		{Vewinger},\ and\ \citenamefont {Weitz}}]{Klaers:2010b}%
	\BibitemOpen
	\bibfield  {author} {\bibinfo {author} {\bibfnamefont {J.}~\bibnamefont
			{Klaers}}, \bibinfo {author} {\bibfnamefont {F.}~\bibnamefont {Vewinger}},\
		and\ \bibinfo {author} {\bibfnamefont {M.}~\bibnamefont {Weitz}},\ }\bibfield
	{title} {\bibinfo {title} {Thermalization of a two-dimensional photonic gas
			in a `white wall' photon box},\ }\href {https://doi.org/10.1038/nphys1680}
	{\bibfield  {journal} {\bibinfo  {journal} {Nat. Phys.}\ }\textbf {\bibinfo
			{volume} {6}},\ \bibinfo {pages} {512} (\bibinfo {year}
		{2010}{\natexlab{a}})}\BibitemShut {NoStop}%
	\bibitem [{\citenamefont {Klaers}\ \emph
		{et~al.}(2010{\natexlab{b}})\citenamefont {Klaers}, \citenamefont {Schmitt},
		\citenamefont {Vewinger},\ and\ \citenamefont {Weitz}}]{Klaers:2010}%
	\BibitemOpen
	\bibfield  {author} {\bibinfo {author} {\bibfnamefont {J.}~\bibnamefont
			{Klaers}}, \bibinfo {author} {\bibfnamefont {J.}~\bibnamefont {Schmitt}},
		\bibinfo {author} {\bibfnamefont {F.}~\bibnamefont {Vewinger}},\ and\
		\bibinfo {author} {\bibfnamefont {M.}~\bibnamefont {Weitz}},\ }\bibfield
	{title} {\bibinfo {title} {{B}ose--{E}instein condensation of photons in an
			optical microcavity},\ }\href {https://doi.org/10.1038/nature09567}
	{\bibfield  {journal} {\bibinfo  {journal} {Nature}\ }\textbf {\bibinfo
			{volume} {468}},\ \bibinfo {pages} {545} (\bibinfo {year}
		{2010}{\natexlab{b}})}\BibitemShut {NoStop}%
	\bibitem [{\citenamefont {Marelic}\ and\ \citenamefont
		{Nyman}(2015)}]{Marelic:2015}%
	\BibitemOpen
	\bibfield  {author} {\bibinfo {author} {\bibfnamefont {J.}~\bibnamefont
			{Marelic}}\ and\ \bibinfo {author} {\bibfnamefont {R.~A.}\ \bibnamefont
			{Nyman}},\ }\bibfield  {title} {\bibinfo {title} {Experimental evidence for
			inhomogeneous pumping and energy-dependent effects in photon
			{B}ose-{E}instein condensation},\ }\href
	{https://doi.org/10.1103/PhysRevA.91.033813} {\bibfield  {journal} {\bibinfo
			{journal} {Phys. Rev. A}\ }\textbf {\bibinfo {volume} {91}},\ \bibinfo
		{pages} {033813} (\bibinfo {year} {2015})}\BibitemShut {NoStop}%
	\bibitem [{\citenamefont {Greveling}\ \emph {et~al.}(2018)\citenamefont
		{Greveling}, \citenamefont {Perrier},\ and\ \citenamefont {van
			Oosten}}]{Greveling:2018}%
	\BibitemOpen
	\bibfield  {author} {\bibinfo {author} {\bibfnamefont {S.}~\bibnamefont
			{Greveling}}, \bibinfo {author} {\bibfnamefont {K.~L.}\ \bibnamefont
			{Perrier}},\ and\ \bibinfo {author} {\bibfnamefont {D.}~\bibnamefont {van
				Oosten}},\ }\bibfield  {title} {\bibinfo {title} {Density distribution of a
			{B}ose-{E}instein condensate of photons in a dye-filled microcavity},\ }\href
	{https://doi.org/10.1103/PhysRevA.98.013810} {\bibfield  {journal} {\bibinfo
			{journal} {Phys. Rev. A}\ }\textbf {\bibinfo {volume} {98}},\ \bibinfo
		{pages} {013810} (\bibinfo {year} {2018})}\BibitemShut {NoStop}%
	\bibitem [{\citenamefont {Vretenar}\ \emph
		{et~al.}(2021{\natexlab{a}})\citenamefont {Vretenar}, \citenamefont
		{Toebes},\ and\ \citenamefont {Klaers}}]{Vretenar:2021a}%
	\BibitemOpen
	\bibfield  {author} {\bibinfo {author} {\bibfnamefont {M.}~\bibnamefont
			{Vretenar}}, \bibinfo {author} {\bibfnamefont {C.}~\bibnamefont {Toebes}},\
		and\ \bibinfo {author} {\bibfnamefont {J.}~\bibnamefont {Klaers}},\
	}\bibfield  {title} {\bibinfo {title} {Modified {B}ose-{E}instein
			condensation in an optical quantum gas},\ }\href
	{https://doi.org/10.1038/s41467-021-26087-0} {\bibfield  {journal} {\bibinfo
			{journal} {Nat. Comm.}\ }\textbf {\bibinfo {volume} {12}},\ \bibinfo {pages}
		{5749} (\bibinfo {year} {2021}{\natexlab{a}})}\BibitemShut {NoStop}%
	\bibitem [{\citenamefont {Schofield}\ \emph {et~al.}(2024)\citenamefont
		{Schofield}, \citenamefont {Fu}, \citenamefont {Clarke}, \citenamefont
		{Farrer}, \citenamefont {Trapalis}, \citenamefont {Dhar}, \citenamefont
		{Mukherjee}, \citenamefont {Severs~Millard}, \citenamefont {Heffernan},
		\citenamefont {Mintert} \emph {et~al.}}]{Schofield:2024}%
	\BibitemOpen
	\bibfield  {author} {\bibinfo {author} {\bibfnamefont {R.~C.}\ \bibnamefont
			{Schofield}}, \bibinfo {author} {\bibfnamefont {M.}~\bibnamefont {Fu}},
		\bibinfo {author} {\bibfnamefont {E.}~\bibnamefont {Clarke}}, \bibinfo
		{author} {\bibfnamefont {I.}~\bibnamefont {Farrer}}, \bibinfo {author}
		{\bibfnamefont {A.}~\bibnamefont {Trapalis}}, \bibinfo {author}
		{\bibfnamefont {H.~S.}\ \bibnamefont {Dhar}}, \bibinfo {author}
		{\bibfnamefont {R.}~\bibnamefont {Mukherjee}}, \bibinfo {author}
		{\bibfnamefont {T.}~\bibnamefont {Severs~Millard}}, \bibinfo {author}
		{\bibfnamefont {J.}~\bibnamefont {Heffernan}}, \bibinfo {author}
		{\bibfnamefont {F.}~\bibnamefont {Mintert}}, \emph {et~al.},\ }\bibfield
	{title} {\bibinfo {title} {Bose--{E}instein condensation of light in a
			semiconductor quantum well microcavity},\ }\href
	{https://doi.org/10.1038/s41566-024-01491-2} {\bibfield  {journal} {\bibinfo
			{journal} {Nat. Photonics}\ }\textbf {\bibinfo {volume} {18}},\ \bibinfo
		{pages} {1083} (\bibinfo {year} {2024})}\BibitemShut {NoStop}%
	\bibitem [{\citenamefont {Pieczarka}\ \emph {et~al.}(2024)\citenamefont
		{Pieczarka}, \citenamefont {G{\k{e}}bski}, \citenamefont {Piasecka},
		\citenamefont {Lott}, \citenamefont {Pelster}, \citenamefont {Wasiak},\ and\
		\citenamefont {Czyszanowski}}]{Pieczarka:2024}%
	\BibitemOpen
	\bibfield  {author} {\bibinfo {author} {\bibfnamefont {M.}~\bibnamefont
			{Pieczarka}}, \bibinfo {author} {\bibfnamefont {M.}~\bibnamefont
			{G{\k{e}}bski}}, \bibinfo {author} {\bibfnamefont {A.~N.}\ \bibnamefont
			{Piasecka}}, \bibinfo {author} {\bibfnamefont {J.~A.}\ \bibnamefont {Lott}},
		\bibinfo {author} {\bibfnamefont {A.}~\bibnamefont {Pelster}}, \bibinfo
		{author} {\bibfnamefont {M.}~\bibnamefont {Wasiak}},\ and\ \bibinfo {author}
		{\bibfnamefont {T.}~\bibnamefont {Czyszanowski}},\ }\bibfield  {title}
	{\bibinfo {title} {Bose--{E}instein condensation of photons in a
			vertical-cavity surface-emitting laser},\ }\href
	{https://doi.org/10.1038/s41566-024-01478-z} {\bibfield  {journal} {\bibinfo
			{journal} {Nat. Photonics}\ }\textbf {\bibinfo {volume} {18}},\ \bibinfo
		{pages} {1090} (\bibinfo {year} {2024})}\BibitemShut {NoStop}%
	\bibitem [{\citenamefont {Kasprzak}\ \emph {et~al.}(2006)\citenamefont
		{Kasprzak}, \citenamefont {Richard}, \citenamefont {Kundermann},
		\citenamefont {Baas}, \citenamefont {Jeambrun}, \citenamefont {Keeling},
		\citenamefont {Marchetti}, \citenamefont {Szymanska}, \citenamefont {Andre},
		\citenamefont {Staehli}, \citenamefont {Savona}, \citenamefont {Littlewood},
		\citenamefont {Deveaud},\ and\ \citenamefont {Dang}}]{Kasprzak:2006}%
	\BibitemOpen
	\bibfield  {author} {\bibinfo {author} {\bibfnamefont {J.}~\bibnamefont
			{Kasprzak}}, \bibinfo {author} {\bibfnamefont {M.}~\bibnamefont {Richard}},
		\bibinfo {author} {\bibfnamefont {S.}~\bibnamefont {Kundermann}}, \bibinfo
		{author} {\bibfnamefont {A.}~\bibnamefont {Baas}}, \bibinfo {author}
		{\bibfnamefont {P.}~\bibnamefont {Jeambrun}}, \bibinfo {author}
		{\bibfnamefont {J.~M.~J.}\ \bibnamefont {Keeling}}, \bibinfo {author}
		{\bibfnamefont {F.~M.}\ \bibnamefont {Marchetti}}, \bibinfo {author}
		{\bibfnamefont {M.~H.}\ \bibnamefont {Szymanska}}, \bibinfo {author}
		{\bibfnamefont {R.}~\bibnamefont {Andre}}, \bibinfo {author} {\bibfnamefont
			{J.~L.}\ \bibnamefont {Staehli}}, \bibinfo {author} {\bibfnamefont
			{V.}~\bibnamefont {Savona}}, \bibinfo {author} {\bibfnamefont {P.~B.}\
			\bibnamefont {Littlewood}}, \bibinfo {author} {\bibfnamefont
			{B.}~\bibnamefont {Deveaud}},\ and\ \bibinfo {author} {\bibfnamefont {L.~S.}\
			\bibnamefont {Dang}},\ }\bibfield  {title} {\bibinfo {title}
		{{B}ose--{E}instein condensation of exciton polaritons},\ }\href
	{https://doi.org/10.1038/nature05131} {\bibfield  {journal} {\bibinfo
			{journal} {Nature}\ }\textbf {\bibinfo {volume} {443}},\ \bibinfo {pages}
		{409} (\bibinfo {year} {2006})}\BibitemShut {NoStop}%
	\bibitem [{\citenamefont {Balili}\ \emph {et~al.}(2007)\citenamefont {Balili},
		\citenamefont {Hartwell}, \citenamefont {Snoke}, \citenamefont {Pfeiffer},\
		and\ \citenamefont {West}}]{Balili:2007}%
	\BibitemOpen
	\bibfield  {author} {\bibinfo {author} {\bibfnamefont {R.}~\bibnamefont
			{Balili}}, \bibinfo {author} {\bibfnamefont {V.}~\bibnamefont {Hartwell}},
		\bibinfo {author} {\bibfnamefont {D.}~\bibnamefont {Snoke}}, \bibinfo
		{author} {\bibfnamefont {L.}~\bibnamefont {Pfeiffer}},\ and\ \bibinfo
		{author} {\bibfnamefont {K.}~\bibnamefont {West}},\ }\bibfield  {title}
	{\bibinfo {title} {Bose-{E}instein condensation of microcavity polaritons in
			a trap},\ }\href {https://doi.org/10.1126/science.1140990} {\bibfield
		{journal} {\bibinfo  {journal} {Science}\ }\textbf {\bibinfo {volume}
			{316}},\ \bibinfo {pages} {1007} (\bibinfo {year} {2007})}\BibitemShut
	{NoStop}%
	\bibitem [{\citenamefont {Deng}\ \emph {et~al.}(2010)\citenamefont {Deng},
		\citenamefont {Haug},\ and\ \citenamefont {Yamamoto}}]{Deng:2010}%
	\BibitemOpen
	\bibfield  {author} {\bibinfo {author} {\bibfnamefont {H.}~\bibnamefont
			{Deng}}, \bibinfo {author} {\bibfnamefont {H.}~\bibnamefont {Haug}},\ and\
		\bibinfo {author} {\bibfnamefont {Y.}~\bibnamefont {Yamamoto}},\ }\bibfield
	{title} {\bibinfo {title} {Exciton-polariton {B}ose-{E}instein
			condensation},\ }\href {https://doi.org/10.1103/RevModPhys.82.1489}
	{\bibfield  {journal} {\bibinfo  {journal} {Rev. Mod. Phys.}\ }\textbf
		{\bibinfo {volume} {82}},\ \bibinfo {pages} {1489} (\bibinfo {year}
		{2010})}\BibitemShut {NoStop}%
	\bibitem [{\citenamefont {Bloch}\ \emph {et~al.}(2022)\citenamefont {Bloch},
		\citenamefont {Carusotto},\ and\ \citenamefont {Wouters}}]{Bloch:2022}%
	\BibitemOpen
	\bibfield  {author} {\bibinfo {author} {\bibfnamefont {J.}~\bibnamefont
			{Bloch}}, \bibinfo {author} {\bibfnamefont {I.}~\bibnamefont {Carusotto}},\
		and\ \bibinfo {author} {\bibfnamefont {M.}~\bibnamefont {Wouters}},\
	}\bibfield  {title} {\bibinfo {title} {Non-equilibrium {B}ose--{E}instein
			condensation in photonic systems},\ }\href
	{https://doi.org/10.1038/s42254-022-00464-0} {\bibfield  {journal} {\bibinfo
			{journal} {Nat. Rev. Phys.}\ }\textbf {\bibinfo {volume} {4}},\ \bibinfo
		{pages} {470–488} (\bibinfo {year} {2022})}\BibitemShut {NoStop}%
	\bibitem [{\citenamefont {Vretenar}\ \emph
		{et~al.}(2021{\natexlab{b}})\citenamefont {Vretenar}, \citenamefont
		{Kassenberg}, \citenamefont {Bissesar}, \citenamefont {Toebes},\ and\
		\citenamefont {Klaers}}]{Vretenar:2021b}%
	\BibitemOpen
	\bibfield  {author} {\bibinfo {author} {\bibfnamefont {M.}~\bibnamefont
			{Vretenar}}, \bibinfo {author} {\bibfnamefont {B.}~\bibnamefont
			{Kassenberg}}, \bibinfo {author} {\bibfnamefont {S.}~\bibnamefont
			{Bissesar}}, \bibinfo {author} {\bibfnamefont {C.}~\bibnamefont {Toebes}},\
		and\ \bibinfo {author} {\bibfnamefont {J.}~\bibnamefont {Klaers}},\
	}\bibfield  {title} {\bibinfo {title} {Controllable {J}osephson junction for
			photon {B}ose-{E}instein condensates},\ }\href
	{https://doi.org/10.1103/PhysRevResearch.3.023167} {\bibfield  {journal}
		{\bibinfo  {journal} {Phys. Rev. Res.}\ }\textbf {\bibinfo {volume} {3}},\
		\bibinfo {pages} {023167} (\bibinfo {year} {2021}{\natexlab{b}})}\BibitemShut
	{NoStop}%
	\bibitem [{\citenamefont {Lagoudakis}\ \emph {et~al.}(2010)\citenamefont
		{Lagoudakis}, \citenamefont {Pietka}, \citenamefont {Wouters}, \citenamefont
		{Andr\'e},\ and\ \citenamefont {Deveaud-Pl\'edran}}]{Lagoudakis:2010}%
	\BibitemOpen
	\bibfield  {author} {\bibinfo {author} {\bibfnamefont {K.~G.}\ \bibnamefont
			{Lagoudakis}}, \bibinfo {author} {\bibfnamefont {B.}~\bibnamefont {Pietka}},
		\bibinfo {author} {\bibfnamefont {M.}~\bibnamefont {Wouters}}, \bibinfo
		{author} {\bibfnamefont {R.}~\bibnamefont {Andr\'e}},\ and\ \bibinfo {author}
		{\bibfnamefont {B.}~\bibnamefont {Deveaud-Pl\'edran}},\ }\bibfield  {title}
	{\bibinfo {title} {Coherent oscillations in an exciton-polariton {J}osephson
			junction},\ }\href {https://doi.org/10.1103/PhysRevLett.105.120403}
	{\bibfield  {journal} {\bibinfo  {journal} {Phys. Rev. Lett.}\ }\textbf
		{\bibinfo {volume} {105}},\ \bibinfo {pages} {120403} (\bibinfo {year}
		{2010})}\BibitemShut {NoStop}%
	\bibitem [{\citenamefont {Abbarchi}\ \emph {et~al.}(2013)\citenamefont
		{Abbarchi}, \citenamefont {Amo}, \citenamefont {Sala}, \citenamefont
		{Solnyshkov}, \citenamefont {Flayac}, \citenamefont {Ferrier}, \citenamefont
		{Sagnes}, \citenamefont {Galopin}, \citenamefont {Lema{\^\i}tre},
		\citenamefont {Malpuech} \emph {et~al.}}]{Abbarchi:2013}%
	\BibitemOpen
	\bibfield  {author} {\bibinfo {author} {\bibfnamefont {M.}~\bibnamefont
			{Abbarchi}}, \bibinfo {author} {\bibfnamefont {A.}~\bibnamefont {Amo}},
		\bibinfo {author} {\bibfnamefont {V.}~\bibnamefont {Sala}}, \bibinfo {author}
		{\bibfnamefont {D.}~\bibnamefont {Solnyshkov}}, \bibinfo {author}
		{\bibfnamefont {H.}~\bibnamefont {Flayac}}, \bibinfo {author} {\bibfnamefont
			{L.}~\bibnamefont {Ferrier}}, \bibinfo {author} {\bibfnamefont
			{I.}~\bibnamefont {Sagnes}}, \bibinfo {author} {\bibfnamefont
			{E.}~\bibnamefont {Galopin}}, \bibinfo {author} {\bibfnamefont
			{A.}~\bibnamefont {Lema{\^\i}tre}}, \bibinfo {author} {\bibfnamefont
			{G.}~\bibnamefont {Malpuech}}, \emph {et~al.},\ }\bibfield  {title} {\bibinfo
		{title} {Macroscopic quantum self-trapping and {J}osephson oscillations of
			exciton polaritons},\ }\href {https://doi.org/10.1038/nphys2609} {\bibfield
		{journal} {\bibinfo  {journal} {Nat. Phys.}\ }\textbf {\bibinfo {volume}
			{9}},\ \bibinfo {pages} {275} (\bibinfo {year} {2013})}\BibitemShut {NoStop}%
	\bibitem [{\citenamefont {Aschieri}\ \emph {et~al.}(2011)\citenamefont
		{Aschieri}, \citenamefont {Garnier}, \citenamefont {Michel}, \citenamefont
		{Doya},\ and\ \citenamefont {Picozzi}}]{Aschieri:2011}%
	\BibitemOpen
	\bibfield  {author} {\bibinfo {author} {\bibfnamefont {P.}~\bibnamefont
			{Aschieri}}, \bibinfo {author} {\bibfnamefont {J.}~\bibnamefont {Garnier}},
		\bibinfo {author} {\bibfnamefont {C.}~\bibnamefont {Michel}}, \bibinfo
		{author} {\bibfnamefont {V.}~\bibnamefont {Doya}},\ and\ \bibinfo {author}
		{\bibfnamefont {A.}~\bibnamefont {Picozzi}},\ }\bibfield  {title} {\bibinfo
		{title} {Condensation and thermalization of classsical optical waves in a
			waveguide},\ }\href {https://doi.org/10.1103/PhysRevA.83.033838} {\bibfield
		{journal} {\bibinfo  {journal} {Phys. Rev. A}\ }\textbf {\bibinfo {volume}
			{83}},\ \bibinfo {pages} {033838} (\bibinfo {year} {2011})}\BibitemShut
	{NoStop}%
	\bibitem [{\citenamefont {Baudin}\ \emph {et~al.}(2020)\citenamefont {Baudin},
		\citenamefont {Fusaro}, \citenamefont {Krupa}, \citenamefont {Garnier},
		\citenamefont {Rica}, \citenamefont {Millot},\ and\ \citenamefont
		{Picozzi}}]{Baudin:2020}%
	\BibitemOpen
	\bibfield  {author} {\bibinfo {author} {\bibfnamefont {K.}~\bibnamefont
			{Baudin}}, \bibinfo {author} {\bibfnamefont {A.}~\bibnamefont {Fusaro}},
		\bibinfo {author} {\bibfnamefont {K.}~\bibnamefont {Krupa}}, \bibinfo
		{author} {\bibfnamefont {J.}~\bibnamefont {Garnier}}, \bibinfo {author}
		{\bibfnamefont {S.}~\bibnamefont {Rica}}, \bibinfo {author} {\bibfnamefont
			{G.}~\bibnamefont {Millot}},\ and\ \bibinfo {author} {\bibfnamefont
			{A.}~\bibnamefont {Picozzi}},\ }\bibfield  {title} {\bibinfo {title}
		{Classical rayleigh-jeans condensation of light waves: Observation and
			thermodynamic characterization},\ }\href
	{https://doi.org/10.1103/PhysRevLett.125.244101} {\bibfield  {journal}
		{\bibinfo  {journal} {Phys. Rev. Lett.}\ }\textbf {\bibinfo {volume} {125}},\
		\bibinfo {pages} {244101} (\bibinfo {year} {2020})}\BibitemShut {NoStop}%
	\bibitem [{\citenamefont {Pourbeyram}\ \emph {et~al.}(2022)\citenamefont
		{Pourbeyram}, \citenamefont {Sidorenko}, \citenamefont {Wu}, \citenamefont
		{Bender}, \citenamefont {Wright}, \citenamefont {Christodoulides},\ and\
		\citenamefont {Wise}}]{Pourbeyram:2022}%
	\BibitemOpen
	\bibfield  {author} {\bibinfo {author} {\bibfnamefont {H.}~\bibnamefont
			{Pourbeyram}}, \bibinfo {author} {\bibfnamefont {P.}~\bibnamefont
			{Sidorenko}}, \bibinfo {author} {\bibfnamefont {F.~O.}\ \bibnamefont {Wu}},
		\bibinfo {author} {\bibfnamefont {N.}~\bibnamefont {Bender}}, \bibinfo
		{author} {\bibfnamefont {L.}~\bibnamefont {Wright}}, \bibinfo {author}
		{\bibfnamefont {D.~N.}\ \bibnamefont {Christodoulides}},\ and\ \bibinfo
		{author} {\bibfnamefont {F.}~\bibnamefont {Wise}},\ }\bibfield  {title}
	{\bibinfo {title} {Direct observations of thermalization to a
			{R}ayleigh--{J}eans distribution in multimode optical fibres},\ }\href
	{https://doi.org/10.1038/s41567-022-01579-y} {\bibfield  {journal} {\bibinfo
			{journal} {Nat. Phys.}\ }\textbf {\bibinfo {volume} {18}},\ \bibinfo {pages}
		{685} (\bibinfo {year} {2022})}\BibitemShut {NoStop}%
	\bibitem [{\citenamefont {Ferraro}\ \emph {et~al.}(2024)\citenamefont
		{Ferraro}, \citenamefont {Mangini}, \citenamefont {Wu}, \citenamefont
		{Zitelli}, \citenamefont {Christodoulides},\ and\ \citenamefont
		{Wabnitz}}]{Ferraro:2024}%
	\BibitemOpen
	\bibfield  {author} {\bibinfo {author} {\bibfnamefont {M.}~\bibnamefont
			{Ferraro}}, \bibinfo {author} {\bibfnamefont {F.}~\bibnamefont {Mangini}},
		\bibinfo {author} {\bibfnamefont {F.~O.}\ \bibnamefont {Wu}}, \bibinfo
		{author} {\bibfnamefont {M.}~\bibnamefont {Zitelli}}, \bibinfo {author}
		{\bibfnamefont {D.~N.}\ \bibnamefont {Christodoulides}},\ and\ \bibinfo
		{author} {\bibfnamefont {S.}~\bibnamefont {Wabnitz}},\ }\bibfield  {title}
	{\bibinfo {title} {Calorimetry of photon gases in nonlinear multimode optical
			fibers},\ }\href {https://doi.org/10.1103/PhysRevX.14.021020} {\bibfield
		{journal} {\bibinfo  {journal} {Phys. Rev. X}\ }\textbf {\bibinfo {volume}
			{14}},\ \bibinfo {pages} {021020} (\bibinfo {year} {2024})}\BibitemShut
	{NoStop}%
	\bibitem [{\citenamefont {Kurtscheid}\ \emph {et~al.}(2019)\citenamefont
		{Kurtscheid}, \citenamefont {Dung}, \citenamefont {Busley}, \citenamefont
		{Vewinger}, \citenamefont {Rosch},\ and\ \citenamefont
		{Weitz}}]{Kurtscheid:2019}%
	\BibitemOpen
	\bibfield  {author} {\bibinfo {author} {\bibfnamefont {C.}~\bibnamefont
			{Kurtscheid}}, \bibinfo {author} {\bibfnamefont {D.}~\bibnamefont {Dung}},
		\bibinfo {author} {\bibfnamefont {E.}~\bibnamefont {Busley}}, \bibinfo
		{author} {\bibfnamefont {F.}~\bibnamefont {Vewinger}}, \bibinfo {author}
		{\bibfnamefont {A.}~\bibnamefont {Rosch}},\ and\ \bibinfo {author}
		{\bibfnamefont {M.}~\bibnamefont {Weitz}},\ }\bibfield  {title} {\bibinfo
		{title} {Thermally condensing photons into a coherently split state of
			light},\ }\href {https://doi.org/10.1126/science.aay1334} {\bibfield
		{journal} {\bibinfo  {journal} {Science}\ }\textbf {\bibinfo {volume}
			{366}},\ \bibinfo {pages} {894} (\bibinfo {year} {2019})}\BibitemShut
	{NoStop}%
	\bibitem [{\citenamefont {Abragam}(1983)}]{Abragam:1983}%
	\BibitemOpen
	\bibfield  {author} {\bibinfo {author} {\bibfnamefont {A.}~\bibnamefont
			{Abragam}},\ }\href
	{https://global.oup.com/academic/product/the-principles-of-nuclear-magnetism-9780198520146?cc=de&lang=en&}
	{\emph {\bibinfo {title} {The Principles of Nuclear Magnetism}}},\
	International series of monographs on physics\ (\bibinfo  {publisher} {Oxford
		University Press},\ \bibinfo {year} {1983})\BibitemShut {NoStop}%
	\bibitem [{\citenamefont {Levitt}(2013)}]{Levitt:2013}%
	\BibitemOpen
	\bibfield  {author} {\bibinfo {author} {\bibfnamefont {M.}~\bibnamefont
			{Levitt}},\ }\href {https://books.google.de/books?id=bysFAa4MPQcC} {\emph
		{\bibinfo {title} {Spin Dynamics: Basics of Nuclear Magnetic Resonance}}}\
	(\bibinfo  {publisher} {Wiley},\ \bibinfo {year} {2013})\BibitemShut
	{NoStop}%
	\bibitem [{\citenamefont {Faghihi}\ \emph {et~al.}(2017)\citenamefont
		{Faghihi}, \citenamefont {Zeinali-Rafsanjani}, \citenamefont
		{Mosleh-Shirazi}, \citenamefont {Saeedi-Moghadam}, \citenamefont {Lotfi},
		\citenamefont {Jalli},\ and\ \citenamefont {Iravani}}]{Faghihi:2017}%
	\BibitemOpen
	\bibfield  {author} {\bibinfo {author} {\bibfnamefont {R.}~\bibnamefont
			{Faghihi}}, \bibinfo {author} {\bibfnamefont {B.}~\bibnamefont
			{Zeinali-Rafsanjani}}, \bibinfo {author} {\bibfnamefont {M.-A.}\ \bibnamefont
			{Mosleh-Shirazi}}, \bibinfo {author} {\bibfnamefont {M.}~\bibnamefont
			{Saeedi-Moghadam}}, \bibinfo {author} {\bibfnamefont {M.}~\bibnamefont
			{Lotfi}}, \bibinfo {author} {\bibfnamefont {R.}~\bibnamefont {Jalli}},\ and\
		\bibinfo {author} {\bibfnamefont {V.}~\bibnamefont {Iravani}},\ }\bibfield
	{title} {\bibinfo {title} {Magnetic resonance spectroscopy and its clinical
			applications: A review},\ }\href
	{https://doi.org/https://doi.org/10.1016/j.jmir.2017.06.004} {\bibfield
		{journal} {\bibinfo  {journal} {J. Med. Imaging Radiat.}\ }\textbf {\bibinfo
			{volume} {48}},\ \bibinfo {pages} {233} (\bibinfo {year} {2017})}\BibitemShut
	{NoStop}%
	\bibitem [{\citenamefont {De~Raedt}\ and\ \citenamefont
		{De~Raedt}(1984)}]{DeRaedt:1984}%
	\BibitemOpen
	\bibfield  {author} {\bibinfo {author} {\bibfnamefont {B.}~\bibnamefont
			{De~Raedt}}\ and\ \bibinfo {author} {\bibfnamefont {H.}~\bibnamefont
			{De~Raedt}},\ }\bibfield  {title} {\bibinfo {title} {Thermodynamics of a
			two-level system coupled to bosons},\ }\href
	{https://doi.org/10.1103/PhysRevB.29.5325} {\bibfield  {journal} {\bibinfo
			{journal} {Phys. Rev. B}\ }\textbf {\bibinfo {volume} {29}},\ \bibinfo
		{pages} {5325} (\bibinfo {year} {1984})}\BibitemShut {NoStop}%
	\bibitem [{\citenamefont {Carmeli}\ and\ \citenamefont
		{Chandler}(1985)}]{Carmeli:1985}%
	\BibitemOpen
	\bibfield  {author} {\bibinfo {author} {\bibfnamefont {B.}~\bibnamefont
			{Carmeli}}\ and\ \bibinfo {author} {\bibfnamefont {D.}~\bibnamefont
			{Chandler}},\ }\bibfield  {title} {\bibinfo {title} {{Effective adiabatic
				approximation for a two level system coupled to a bath}},\ }\href
	{https://doi.org/10.1063/1.448942} {\bibfield  {journal} {\bibinfo  {journal}
			{J. Chem. Phys.}\ }\textbf {\bibinfo {volume} {82}},\ \bibinfo {pages} {3400}
		(\bibinfo {year} {1985})}\BibitemShut {NoStop}%
	\bibitem [{\citenamefont {Weiss}(2008)}]{Weiss:2008}%
	\BibitemOpen
	\bibfield  {author} {\bibinfo {author} {\bibfnamefont {U.}~\bibnamefont
			{Weiss}},\ }\href {https://books.google.de/books?id=mGVhDQAAQBAJ} {\emph
		{\bibinfo {title} {Quantum Dissipative Systems}}},\ Series in modern
	condensed matter physics\ (\bibinfo  {publisher} {World Scientific},\
	\bibinfo {year} {2008})\BibitemShut {NoStop}%
	\bibitem [{\citenamefont {Kockum}\ and\ \citenamefont
		{Nori}(2019)}]{Kockum:2019}%
	\BibitemOpen
	\bibfield  {author} {\bibinfo {author} {\bibfnamefont {A.~F.}\ \bibnamefont
			{Kockum}}\ and\ \bibinfo {author} {\bibfnamefont {F.}~\bibnamefont {Nori}},\
	}\bibinfo {title} {Quantum bits with {J}osephson junctions},\ in\ \href
	{https://doi.org/10.1007/978-3-030-20726-7_17} {\emph {\bibinfo {booktitle}
			{Fundamentals and Frontiers of the Josephson Effect}}},\ \bibinfo {editor}
	{edited by\ \bibinfo {editor} {\bibfnamefont {F.}~\bibnamefont {Tafuri}}}\
	(\bibinfo  {publisher} {Springer International Publishing},\ \bibinfo
	{address} {Cham},\ \bibinfo {year} {2019})\ pp.\ \bibinfo {pages}
	{703--741}\BibitemShut {NoStop}%
	\bibitem [{\citenamefont {Sachdev}(2011)}]{Sachdev:2011}%
	\BibitemOpen
	\bibfield  {author} {\bibinfo {author} {\bibfnamefont {S.}~\bibnamefont
			{Sachdev}},\ }\href@noop {} {\emph {\bibinfo {title} {Quantum Phase
				Transitions}}},\ \bibinfo {edition} {2nd}\ ed.\ (\bibinfo  {publisher}
	{Cambridge University Press},\ \bibinfo {year} {2011})\BibitemShut {NoStop}%
	\bibitem [{\citenamefont {Albiez}\ \emph {et~al.}(2005)\citenamefont {Albiez},
		\citenamefont {Gati}, \citenamefont {F{\"o}lling}, \citenamefont {Hunsmann},
		\citenamefont {Cristiani},\ and\ \citenamefont {Oberthaler}}]{Albiez:2005}%
	\BibitemOpen
	\bibfield  {author} {\bibinfo {author} {\bibfnamefont {M.}~\bibnamefont
			{Albiez}}, \bibinfo {author} {\bibfnamefont {R.}~\bibnamefont {Gati}},
		\bibinfo {author} {\bibfnamefont {J.}~\bibnamefont {F{\"o}lling}}, \bibinfo
		{author} {\bibfnamefont {S.}~\bibnamefont {Hunsmann}}, \bibinfo {author}
		{\bibfnamefont {M.}~\bibnamefont {Cristiani}},\ and\ \bibinfo {author}
		{\bibfnamefont {M.}~\bibnamefont {Oberthaler}},\ }\bibfield  {title}
	{\bibinfo {title} {{Direct observation of tunneling and nonlinear
				self-trapping in a single bosonic {J}osephson junction}},\ }\href
	{https://doi.org/10.1103/PhysRevLett.95.010402} {\bibfield  {journal}
		{\bibinfo  {journal} {Phys. Rev. Lett.}\ }\textbf {\bibinfo {volume} {95}},\
		\bibinfo {pages} {10402} (\bibinfo {year} {2005})}\BibitemShut {NoStop}%
	\bibitem [{\citenamefont {Esteve}\ \emph {et~al.}(2008)\citenamefont {Esteve},
		\citenamefont {Gross}, \citenamefont {A.Weller}, \citenamefont
		{S.Giovanazzi},\ and\ \citenamefont {Oberthaler}}]{Esteve:2008}%
	\BibitemOpen
	\bibfield  {author} {\bibinfo {author} {\bibfnamefont {J.}~\bibnamefont
			{Esteve}}, \bibinfo {author} {\bibfnamefont {C.}~\bibnamefont {Gross}},
		\bibinfo {author} {\bibnamefont {A.Weller}}, \bibinfo {author} {\bibnamefont
			{S.Giovanazzi}},\ and\ \bibinfo {author} {\bibfnamefont {M.~K.}\ \bibnamefont
			{Oberthaler}},\ }\bibfield  {title} {\bibinfo {title} {Squeezing and
			entanglement in a {B}ose--{E}instein condensate},\ }\href
	{https://doi.org/10.1038/nature07332} {\bibfield  {journal} {\bibinfo
			{journal} {Nature}\ }\textbf {\bibinfo {volume} {455}},\ \bibinfo {pages}
		{1216} (\bibinfo {year} {2008})}\BibitemShut {NoStop}%
	\bibitem [{\citenamefont {Kurtscheid}\ \emph {et~al.}(2020)\citenamefont
		{Kurtscheid}, \citenamefont {Dung}, \citenamefont {Redmann}, \citenamefont
		{Busley}, \citenamefont {Klaers}, \citenamefont {Vewinger}, \citenamefont
		{Schmitt},\ and\ \citenamefont {Weitz}}]{Kurtscheid:2020}%
	\BibitemOpen
	\bibfield  {author} {\bibinfo {author} {\bibfnamefont {C.}~\bibnamefont
			{Kurtscheid}}, \bibinfo {author} {\bibfnamefont {D.}~\bibnamefont {Dung}},
		\bibinfo {author} {\bibfnamefont {A.}~\bibnamefont {Redmann}}, \bibinfo
		{author} {\bibfnamefont {E.}~\bibnamefont {Busley}}, \bibinfo {author}
		{\bibfnamefont {J.}~\bibnamefont {Klaers}}, \bibinfo {author} {\bibfnamefont
			{F.}~\bibnamefont {Vewinger}}, \bibinfo {author} {\bibfnamefont
			{J.}~\bibnamefont {Schmitt}},\ and\ \bibinfo {author} {\bibfnamefont
			{M.}~\bibnamefont {Weitz}},\ }\bibfield  {title} {\bibinfo {title} {Realizing
			arbitrary trapping potentials for light via direct laser writing of mirror
			surface profiles},\ }\href {https://doi.org/10.1209/0295-5075/130/54001}
	{\bibfield  {journal} {\bibinfo  {journal} {{EPL} (Europhys. Lett.)}\
		}\textbf {\bibinfo {volume} {130}},\ \bibinfo {pages} {54001} (\bibinfo
		{year} {2020})}\BibitemShut {NoStop}%
	\bibitem [{\citenamefont {Hesten}\ \emph {et~al.}(2018)\citenamefont {Hesten},
		\citenamefont {Nyman},\ and\ \citenamefont {Mintert}}]{Hesten:2018}%
	\BibitemOpen
	\bibfield  {author} {\bibinfo {author} {\bibfnamefont {H.~J.}\ \bibnamefont
			{Hesten}}, \bibinfo {author} {\bibfnamefont {R.~A.}\ \bibnamefont {Nyman}},\
		and\ \bibinfo {author} {\bibfnamefont {F.}~\bibnamefont {Mintert}},\
	}\bibfield  {title} {\bibinfo {title} {Decondensation in nonequilibrium
			photonic condensates: When less is more},\ }\href
	{https://doi.org/10.1103/PhysRevLett.120.040601} {\bibfield  {journal}
		{\bibinfo  {journal} {Phys. Rev. Lett.}\ }\textbf {\bibinfo {volume} {120}},\
		\bibinfo {pages} {040601} (\bibinfo {year} {2018})}\BibitemShut {NoStop}%
	\bibitem [{\citenamefont {Siegman}(1986)}]{Siegman:1986}%
	\BibitemOpen
	\bibfield  {author} {\bibinfo {author} {\bibfnamefont {A.}~\bibnamefont
			{Siegman}},\ }\href {https://books.google.de/books?id=1BZVwUZLTkAC} {\emph
		{\bibinfo {title} {Lasers}}}\ (\bibinfo  {publisher} {University Science
		Books},\ \bibinfo {address} {Sausalito},\ \bibinfo {year} {1986})\BibitemShut
	{NoStop}%
	\bibitem [{\citenamefont {Klaers}\ \emph {et~al.}(2012)\citenamefont {Klaers},
		\citenamefont {Schmitt}, \citenamefont {Damm}, \citenamefont {Vewinger},\
		and\ \citenamefont {Weitz}}]{Klaers:2012}%
	\BibitemOpen
	\bibfield  {author} {\bibinfo {author} {\bibfnamefont {J.}~\bibnamefont
			{Klaers}}, \bibinfo {author} {\bibfnamefont {J.}~\bibnamefont {Schmitt}},
		\bibinfo {author} {\bibfnamefont {T.}~\bibnamefont {Damm}}, \bibinfo {author}
		{\bibfnamefont {F.}~\bibnamefont {Vewinger}},\ and\ \bibinfo {author}
		{\bibfnamefont {M.}~\bibnamefont {Weitz}},\ }\bibfield  {title} {\bibinfo
		{title} {Statistical physics of {B}ose-{E}instein-condensed light in a dye
			microcavity},\ }\href {https://doi.org/10.1103/PhysRevLett.108.160403}
	{\bibfield  {journal} {\bibinfo  {journal} {Phys. Rev. Lett.}\ }\textbf
		{\bibinfo {volume} {108}},\ \bibinfo {pages} {160403} (\bibinfo {year}
		{2012})}\BibitemShut {NoStop}%
	\bibitem [{\citenamefont {Schmitt}\ \emph {et~al.}(2015)\citenamefont
		{Schmitt}, \citenamefont {Damm}, \citenamefont {Dung}, \citenamefont
		{Vewinger}, \citenamefont {Klaers},\ and\ \citenamefont
		{Weitz}}]{Schmitt:2015}%
	\BibitemOpen
	\bibfield  {author} {\bibinfo {author} {\bibfnamefont {J.}~\bibnamefont
			{Schmitt}}, \bibinfo {author} {\bibfnamefont {T.}~\bibnamefont {Damm}},
		\bibinfo {author} {\bibfnamefont {D.}~\bibnamefont {Dung}}, \bibinfo {author}
		{\bibfnamefont {F.}~\bibnamefont {Vewinger}}, \bibinfo {author}
		{\bibfnamefont {J.}~\bibnamefont {Klaers}},\ and\ \bibinfo {author}
		{\bibfnamefont {M.}~\bibnamefont {Weitz}},\ }\bibfield  {title} {\bibinfo
		{title} {Thermalization kinetics of light: From laser dynamics to equilibrium
			condensation of photons},\ }\href
	{https://doi.org/10.1103/PhysRevA.92.011602} {\bibfield  {journal} {\bibinfo
			{journal} {Phys. Rev. A}\ }\textbf {\bibinfo {volume} {92}},\ \bibinfo
		{pages} {011602} (\bibinfo {year} {2015})}\BibitemShut {NoStop}%
	\bibitem [{\citenamefont {Xue}\ \emph {et~al.}(2021)\citenamefont {Xue},
		\citenamefont {Chestnov}, \citenamefont {Sedov}, \citenamefont {Kiktenko},
		\citenamefont {Fedorov}, \citenamefont {Schumacher}, \citenamefont {Ma},\
		and\ \citenamefont {Kavokin}}]{Xue:2021}%
	\BibitemOpen
	\bibfield  {author} {\bibinfo {author} {\bibfnamefont {Y.}~\bibnamefont
			{Xue}}, \bibinfo {author} {\bibfnamefont {I.}~\bibnamefont {Chestnov}},
		\bibinfo {author} {\bibfnamefont {E.}~\bibnamefont {Sedov}}, \bibinfo
		{author} {\bibfnamefont {E.}~\bibnamefont {Kiktenko}}, \bibinfo {author}
		{\bibfnamefont {A.~K.}\ \bibnamefont {Fedorov}}, \bibinfo {author}
		{\bibfnamefont {S.}~\bibnamefont {Schumacher}}, \bibinfo {author}
		{\bibfnamefont {X.}~\bibnamefont {Ma}},\ and\ \bibinfo {author}
		{\bibfnamefont {A.}~\bibnamefont {Kavokin}},\ }\bibfield  {title} {\bibinfo
		{title} {Split-ring polariton condensates as macroscopic two-level quantum
			systems},\ }\href {https://doi.org/10.1103/PhysRevResearch.3.013099}
	{\bibfield  {journal} {\bibinfo  {journal} {Phys. Rev. Res.}\ }\textbf
		{\bibinfo {volume} {3}},\ \bibinfo {pages} {013099} (\bibinfo {year}
		{2021})}\BibitemShut {NoStop}%
	\bibitem [{\citenamefont {Barrat}\ \emph {et~al.}(2024)\citenamefont {Barrat},
		\citenamefont {Tzortzakakis}, \citenamefont {Niu}, \citenamefont {Zhou},
		\citenamefont {Paschos}, \citenamefont {Petrosyan},\ and\ \citenamefont
		{Savvidis}}]{Barrat:2024}%
	\BibitemOpen
	\bibfield  {author} {\bibinfo {author} {\bibfnamefont {J.}~\bibnamefont
			{Barrat}}, \bibinfo {author} {\bibfnamefont {A.~F.}\ \bibnamefont
			{Tzortzakakis}}, \bibinfo {author} {\bibfnamefont {M.}~\bibnamefont {Niu}},
		\bibinfo {author} {\bibfnamefont {X.}~\bibnamefont {Zhou}}, \bibinfo {author}
		{\bibfnamefont {G.~G.}\ \bibnamefont {Paschos}}, \bibinfo {author}
		{\bibfnamefont {D.}~\bibnamefont {Petrosyan}},\ and\ \bibinfo {author}
		{\bibfnamefont {P.~G.}\ \bibnamefont {Savvidis}},\ }\bibfield  {title}
	{\bibinfo {title} {Qubit analog with polariton superfluid in an annular
			trap},\ }\href {https://doi.org/10.1126/sciadv.ado4042} {\bibfield  {journal}
		{\bibinfo  {journal} {Science Advances}\ }\textbf {\bibinfo {volume} {10}},\
		\bibinfo {pages} {eado4042} (\bibinfo {year} {2024})}\BibitemShut {NoStop}%
	\bibitem [{\citenamefont {Bennett}\ \emph {et~al.}(2020)\citenamefont
		{Bennett}, \citenamefont {Steinbrecht}, \citenamefont {Gorbachev},\ and\
		\citenamefont {Buhmann}}]{Bennet:2020}%
	\BibitemOpen
	\bibfield  {author} {\bibinfo {author} {\bibfnamefont {R.}~\bibnamefont
			{Bennett}}, \bibinfo {author} {\bibfnamefont {D.}~\bibnamefont
			{Steinbrecht}}, \bibinfo {author} {\bibfnamefont {Y.}~\bibnamefont
			{Gorbachev}},\ and\ \bibinfo {author} {\bibfnamefont {S.~Y.}\ \bibnamefont
			{Buhmann}},\ }\bibfield  {title} {\bibinfo {title} {Symmetry breaking in a
			condensate of light and its use as a quantum sensor},\ }\href
	{https://doi.org/10.1103/PhysRevApplied.13.044031} {\bibfield  {journal}
		{\bibinfo  {journal} {Phys. Rev. Appl.}\ }\textbf {\bibinfo {volume} {13}},\
		\bibinfo {pages} {044031} (\bibinfo {year} {2020})}\BibitemShut {NoStop}%
	\bibitem [{\citenamefont {Da~Lio}\ \emph {et~al.}(2019)\citenamefont {Da~Lio},
		\citenamefont {Bacco}, \citenamefont {Cozzolino}, \citenamefont {Ding},
		\citenamefont {Dalgaard}, \citenamefont {Rottwitt},\ and\ \citenamefont
		{Oxenløwe}}]{Da:2019}%
	\BibitemOpen
	\bibfield  {author} {\bibinfo {author} {\bibfnamefont {B.}~\bibnamefont
			{Da~Lio}}, \bibinfo {author} {\bibfnamefont {D.}~\bibnamefont {Bacco}},
		\bibinfo {author} {\bibfnamefont {D.}~\bibnamefont {Cozzolino}}, \bibinfo
		{author} {\bibfnamefont {Y.}~\bibnamefont {Ding}}, \bibinfo {author}
		{\bibfnamefont {K.}~\bibnamefont {Dalgaard}}, \bibinfo {author}
		{\bibfnamefont {K.}~\bibnamefont {Rottwitt}},\ and\ \bibinfo {author}
		{\bibfnamefont {L.~K.}\ \bibnamefont {Oxenløwe}},\ }\bibfield  {title}
	{\bibinfo {title} {Experimental demonstration of the dpts qkd protocol over a
			170 km fiber link},\ }\href {https://doi.org/10.1063/1.5049659} {\bibfield
		{journal} {\bibinfo  {journal} {Applied Physics Letters}\ }\textbf {\bibinfo
			{volume} {114}},\ \bibinfo {pages} {011101} (\bibinfo {year}
		{2019})}\BibitemShut {NoStop}%
	\bibitem [{\citenamefont {Lenzini}\ \emph {et~al.}(2018)\citenamefont
		{Lenzini}, \citenamefont {Janousek}, \citenamefont {Thearle}, \citenamefont
		{Villa}, \citenamefont {Haylock}, \citenamefont {Kasture}, \citenamefont
		{Cui}, \citenamefont {Phan}, \citenamefont {Dao}, \citenamefont {Yonezawa},
		\citenamefont {Lam}, \citenamefont {Huntington},\ and\ \citenamefont
		{Lobino}}]{Lenzini:2018}%
	\BibitemOpen
	\bibfield  {author} {\bibinfo {author} {\bibfnamefont {F.}~\bibnamefont
			{Lenzini}}, \bibinfo {author} {\bibfnamefont {J.}~\bibnamefont {Janousek}},
		\bibinfo {author} {\bibfnamefont {O.}~\bibnamefont {Thearle}}, \bibinfo
		{author} {\bibfnamefont {M.}~\bibnamefont {Villa}}, \bibinfo {author}
		{\bibfnamefont {B.}~\bibnamefont {Haylock}}, \bibinfo {author} {\bibfnamefont
			{S.}~\bibnamefont {Kasture}}, \bibinfo {author} {\bibfnamefont
			{L.}~\bibnamefont {Cui}}, \bibinfo {author} {\bibfnamefont {H.-P.}\
			\bibnamefont {Phan}}, \bibinfo {author} {\bibfnamefont {D.~V.}\ \bibnamefont
			{Dao}}, \bibinfo {author} {\bibfnamefont {H.}~\bibnamefont {Yonezawa}},
		\bibinfo {author} {\bibfnamefont {P.~K.}\ \bibnamefont {Lam}}, \bibinfo
		{author} {\bibfnamefont {E.~H.}\ \bibnamefont {Huntington}},\ and\ \bibinfo
		{author} {\bibfnamefont {M.}~\bibnamefont {Lobino}},\ }\bibfield  {title}
	{\bibinfo {title} {Integrated photonic platform for quantum information with
			continuous variables},\ }\href {https://doi.org/10.1126/sciadv.aat9331}
	{\bibfield  {journal} {\bibinfo  {journal} {Science Advances}\ }\textbf
		{\bibinfo {volume} {4}},\ \bibinfo {pages} {eaat9331} (\bibinfo {year}
		{2018})}\BibitemShut {NoStop}%
	\bibitem [{\citenamefont {Linslal}\ \emph {et~al.}(2022)\citenamefont
		{Linslal}, \citenamefont {Ayyaswamy}, \citenamefont {Maji}, \citenamefont
		{Sooraj}, \citenamefont {Dixit}, \citenamefont {Venkitesh},\ and\
		\citenamefont {Srinivasan}}]{Linslal:2022}%
	\BibitemOpen
	\bibfield  {author} {\bibinfo {author} {\bibfnamefont {C.~L.}\ \bibnamefont
			{Linslal}}, \bibinfo {author} {\bibfnamefont {P.}~\bibnamefont {Ayyaswamy}},
		\bibinfo {author} {\bibfnamefont {S.}~\bibnamefont {Maji}}, \bibinfo {author}
		{\bibfnamefont {M.~S.}\ \bibnamefont {Sooraj}}, \bibinfo {author}
		{\bibfnamefont {A.}~\bibnamefont {Dixit}}, \bibinfo {author} {\bibfnamefont
			{D.}~\bibnamefont {Venkitesh}},\ and\ \bibinfo {author} {\bibfnamefont
			{B.}~\bibnamefont {Srinivasan}},\ }\bibfield  {title} {\bibinfo {title}
		{Challenges in coherent beam combining of high power fiber amplifiers: a
			review},\ }\href {https://doi.org/10.1007/s41683-022-00099-4} {\bibfield
		{journal} {\bibinfo  {journal} {ISSS Journal of Micro and Smart Systems}\
		}\textbf {\bibinfo {volume} {11}},\ \bibinfo {pages} {277} (\bibinfo {year}
		{2022})}\BibitemShut {NoStop}%
	\bibitem [{\citenamefont {Toledo~Tude}\ and\ \citenamefont
		{Eastham}(2024)}]{Tude:2024}%
	\BibitemOpen
	\bibfield  {author} {\bibinfo {author} {\bibfnamefont {L.}~\bibnamefont
			{Toledo~Tude}}\ and\ \bibinfo {author} {\bibfnamefont {P.~R.}\ \bibnamefont
			{Eastham}},\ }\bibfield  {title} {\bibinfo {title} {Quantum thermodynamics of
			driven-dissipative condensates},\ }\href {https://doi.org/10.1063/5.0208352}
	{\bibfield  {journal} {\bibinfo  {journal} {APL Quantum}\ }\textbf {\bibinfo
			{volume} {1}},\ \bibinfo {pages} {036108} (\bibinfo {year}
		{2024})}\BibitemShut {NoStop}%
	\bibitem [{\citenamefont {Williamson}\ \emph {et~al.}(2024)\citenamefont
		{Williamson}, \citenamefont {Cerisola}, \citenamefont {Anders},\ and\
		\citenamefont {Davis}}]{Williamson:2024}%
	\BibitemOpen
	\bibfield  {author} {\bibinfo {author} {\bibfnamefont {L.~A.}\ \bibnamefont
			{Williamson}}, \bibinfo {author} {\bibfnamefont {F.}~\bibnamefont
			{Cerisola}}, \bibinfo {author} {\bibfnamefont {J.}~\bibnamefont {Anders}},\
		and\ \bibinfo {author} {\bibfnamefont {M.~J.}\ \bibnamefont {Davis}},\
	}\bibfield  {title} {\bibinfo {title} {Extracting work from coherence in a
			two-mode bose–einstein condensate},\ }\href
	{https://doi.org/10.1088/2058-9565/ad8fc9} {\bibfield  {journal} {\bibinfo
			{journal} {Quantum Science and Technology}\ }\textbf {\bibinfo {volume}
			{10}},\ \bibinfo {pages} {015040} (\bibinfo {year} {2024})}\BibitemShut
	{NoStop}%
	\bibitem [{\citenamefont {44}}]{Zenodo:2025}%
	\BibitemOpen
	\href {https://doi.org/10.5281/zenodo.17293414}
	{\bibfield  {journal} {\bibinfo  {journal} {https://doi.org/10.5281/zenodo.17293414}}}\BibitemShut {NoStop}%
	\bibitem [{\citenamefont {Vretenar}\ \emph {et~al.}(2023)\citenamefont
		{Vretenar}, \citenamefont {Puplauskis},\ and\ \citenamefont
		{Klaers}}]{Vretenar:2023}%
	\BibitemOpen
	\bibfield  {author} {\bibinfo {author} {\bibfnamefont {M.}~\bibnamefont
			{Vretenar}}, \bibinfo {author} {\bibfnamefont {M.}~\bibnamefont
			{Puplauskis}},\ and\ \bibinfo {author} {\bibfnamefont {J.}~\bibnamefont
			{Klaers}},\ }\bibfield  {title} {\bibinfo {title} {Mirror surface
			nanostructuring via laser direct writing—characterization and physical
			origins},\ }\href {https://doi.org/https://doi.org/10.1002/adom.202202820}
	{\bibfield  {journal} {\bibinfo  {journal} {Advanced Optical Materials}\
		}\textbf {\bibinfo {volume} {11}},\ \bibinfo {pages} {2202820} (\bibinfo
		{year} {2023})}\BibitemShut {NoStop}%
	\bibitem [{\citenamefont {Kirton}\ and\ \citenamefont
		{Keeling}(2013)}]{Kirton:2013}%
	\BibitemOpen
	\bibfield  {author} {\bibinfo {author} {\bibfnamefont {P.}~\bibnamefont
			{Kirton}}\ and\ \bibinfo {author} {\bibfnamefont {J.}~\bibnamefont
			{Keeling}},\ }\bibfield  {title} {\bibinfo {title} {Nonequilibrium model of
			photon condensation},\ }\href
	{https://doi.org/10.1103/PhysRevLett.111.100404} {\bibfield  {journal}
		{\bibinfo  {journal} {Phys. Rev. Lett.}\ }\textbf {\bibinfo {volume} {111}},\
		\bibinfo {pages} {100404} (\bibinfo {year} {2013})}\BibitemShut {NoStop}%
	\bibitem [{\citenamefont {Lakowicz}(2006)}]{Lakowicz:2006}%
	\BibitemOpen
	\bibfield  {author} {\bibinfo {author} {\bibfnamefont {J.}~\bibnamefont
			{Lakowicz}},\ }\href {https://doi.org/10.1007/978-0-387-46312-4} {\emph
		{\bibinfo {title} {Principles of Fluorescence Spectroscopy}}}\ (\bibinfo
	{publisher} {Springer New York, NY},\ \bibinfo {year} {2006})\BibitemShut
	{NoStop}%
	\bibitem [{\citenamefont {Keeling}\ and\ \citenamefont
		{Kirton}(2016)}]{Keeling:2016}%
	\BibitemOpen
	\bibfield  {author} {\bibinfo {author} {\bibfnamefont {J.}~\bibnamefont
			{Keeling}}\ and\ \bibinfo {author} {\bibfnamefont {P.}~\bibnamefont
			{Kirton}},\ }\bibfield  {title} {\bibinfo {title} {Spatial dynamics,
			thermalization, and gain clamping in a photon condensate},\ }\href
	{https://doi.org/10.1103/PhysRevA.93.013829} {\bibfield  {journal} {\bibinfo
			{journal} {Phys. Rev. A}\ }\textbf {\bibinfo {volume} {93}},\ \bibinfo
		{pages} {013829} (\bibinfo {year} {2016})}\BibitemShut {NoStop}%
\end{thebibliography}
\end{document}